\newcommand{\Be}{$^9$Be$^+$~}

\newcommand{\tE}{$\mathcal{E}$}
\newcommand{\tC}{$\mathcal{C}$}
\newcommand{\tF}{$\mathcal{F}$}
\newcommand{\tV}{$\mathcal{V}$}
\newcommand{\tECE}{$\mathcal{E}$-$\mathcal{C}$-$\mathcal{E}$}
\newcommand{\tECFCE}{$\mathcal{E}$-$\mathcal{C}$-$\mathcal{F}$-$\mathcal{C}$-$\mathcal{E}$}
\newcommand{\tECVCE}{$\mathcal{E}$-$\mathcal{C}$-$\mathcal{V}$-$\mathcal{C}$-$\mathcal{E}$}

\newcommand{\OmegaRF}{\Omega_{\mathrm{rf}}}
\newcommand{\OmegaDAC}{R_{\mathrm{DAC}}}

\newcommand{\ket}[1]{\left| #1 \right>}

\newcommand{\abs}[1]{\left| #1 \right|}

\documentclass[twocolumn,pra,showpacs]{revtex4}

\usepackage[pdftex]{graphicx,color}
\usepackage{amsmath}
\usepackage{amssymb}

\begin{document}

\title{Improved high-fidelity transport of trapped-ion qubits through a
multi-dimensional array}

\author{R.B. Blakestad}
	\thanks{Present address: Joint Quantum Institute, NIST and University of
	Maryland, Gaithersburg, MD 20899, USA}
\author{C. Ospelkaus}
	\thanks{Present address: QUEST, Leibniz Universit\"{a}t Hannover,
	30167 Hannover and PTB, 38116 Braunschweig, Germany}
\author{A.P. VanDevender}
	\thanks{Present address: Halcyon Molecular, Redwood City, CA 94063, USA}
\author{J.H.~Wesenberg}
	\thanks{Present address: Centre for Quantum Technologies,
	National University of Singapore, Singapore 117543, Singapore}
\author{M.J. Biercuk}
	\thanks{Present address: School of Physics, The University of Sydney, NSW 2006
	Australia.}
\author{D. Leibfried}
\author{D.J. Wineland}

\affiliation{National Institute of Standards and Technology, 325 Broadway, Boulder,
Colorado 80305, USA }

\date{\today}

\begin{abstract}
We have demonstrated transport of \Be ions through a 2D Paul-trap array that
incorporates an X-junction, while maintaining the ions near the motional
ground-state of the confining potential well. We expand on the first report of
the experiment~\cite{blakestad2009a}, including a detailed discussion of how
the transport potentials were calculated. Two main mechanisms that caused
motional excitation during transport are explained, along with the methods used
to mitigate such excitation. We reduced the motional excitation below the results
in Ref.~\cite{blakestad2009a} by a factor of approximately 50. The effect of a
mu-metal shield on qubit coherence is also reported. Finally, we examined a
method for exchanging energy between multiple motional modes on the few-quanta level,
which could be useful for cooling motional modes without directly accessing the
modes with lasers. These results establish how trapped ions can be transported in
a large-scale quantum processor with high fidelity.
\end{abstract}
 \pacs{37.10.Ty, 37.10.Rs, 03.67.Lx}
\maketitle


\section{Introduction}\label{sec:Introduction}

The reliable transport of quantum information will enable operations between any
arbitrarily selected qubits in a quantum processor and is essential to realize
efficient, large-scale quantum information processing (QIP). Trapped ions are a
promising system in which to study QIP~\cite{blatt2008a, monroe2008a,
haffner2008a}, and several approaches to achieving reliable information transport
have been proposed~\cite{cirac2000a, blatt2008a, monroe2008a, haffner2008a,
bible, kielpinski2002a, duan2010a, lin2009a}. In most demonstrated entangling
gate operations that use ions, qubits stored in the internal atomic states of
ions are entangled by coupling the internal states with a single shared motional
mode through a laser-induced interaction~\cite{cirac1995a, blatt2008a,
monroe2008a, haffner2008a}. However, as the number of ions grows large ($>10$),
it becomes difficult to isolate a single mode during gate
operations~\cite{hughes1996a, bible}. One way around this issue is to distribute
the ions over an array of harmonic potentials, where the number of ions in each
trapping potential can remain small. The potentials can be adjusted temporally to
transport the ions throughout the array and combine selected ions into a
particular harmonic potential. Once combined, gate operations can be performed on
the selected ions by use of a local shared mode of motion~\cite{bible,
kielpinski2002a}.

Initial demonstrations of such distributed architectures have incorporated simple
linear arrays~\cite{rowe2002a, barrett2004a, huber2008a, home2009a, eble2009a},
where all ions are confined in potential minima on a line along an axis of the
trap. The order of ions within the linear array can even be
changed~\cite{splatt2009a}. However, multidimensional
arrays~\cite{kielpinski2002a, bible} provide the greatest flexibility in ion-trap
processor architectures, and permit more efficient reordering of ion strings for
deterministic gate operations.  The key technical element that must be realized
towards this end is the two-dimensional junction, which consists of multiple
intersecting linear arrays. The potentials in a junction are more complicated
than those in a linear array, making transport through a junction challenging.
Since the fidelity of the gates is highest if the ions are near their motional
ground state, it is important that transport through such arrays be performed
reliably and with minimal excitation of the ion's motion in its local trapping
potential. If multiple transports are needed, each transport should contribute
well under a single quantum of motional excitation, though sympathetic cooling
can be used to remove excess motional energy, at the cost of increased experiment
duration (and accompanying decoherence)~\cite{kielpinski2002a}. For simple linear
arrays, reliable transport with little motional excitation has been
demonstrated~\cite{rowe2002a, barrett2004a, home2009a}.

To date, transport through a T-junction~\cite{hensinger2006a}, an
X-junction~\cite{blakestad2009a} and surface-electrode
Y-junctions~\cite{amini2010a, moehring2011a} have been demonstrated. However,
such transport has not yet been demonstrated with sufficiently-low motional
excitation (at or below a single quantum). Using the apparatus in
Ref.~\cite{blakestad2009a}, we have now realized highly reliable transport
through an X-junction with excitation of less than one quantum of motion per
transport, a decrease of approximately  50 compared to the results in
Ref.~\cite{blakestad2009a}. This has allowed us to observe a process where energy
can be exchanged between motional modes in certain situations, and demonstrates
motional control over the ions at the single-quantum level.  The paper is
organized as follows: we begin in Sec.~\ref{sec:Trap} with a description of the
X-junction trap array used for transport. Section~\ref{sec:Waveforms} lays out
the procedure for calculating the time-dependent trapping potentials that
transport the ion. A description of the basic transport experiment is given in
Sec.~\ref{sec:Experiment}. Section~\ref{sec:HeatingSources} covers the various
mechanisms that excite the ion's motion during transport, as well as the
filtering techniques used to mitigate those excitations. This understanding of
the noise sources, and the subsequent improved filtering techniques, allowed the
reduction in motional excitation relative to Ref.~\cite{blakestad2009a}. To
mitigate the effects of magnetic field fluctuations on qubit decoherence, a
mu-metal shield and field-coil current stabilization were used, which is
explained in Sec.~\ref{sec:Shield}. Finally, in Sec.~\ref{sec:Exchange}, we
discuss a procedure for swapping motional energy between motional modes at the
center of the junction array. This swapping process can potentially be used to
laser-cool multiple modes of motion without the need for a direct interaction
between the cooling laser and every motional mode.

\section{X-junction Array}\label{sec:Trap}

\begin{figure}
	\includegraphics{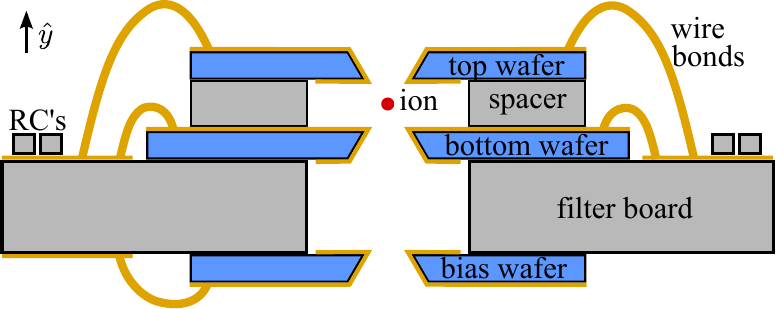}
	\caption{A cross sectional view (not to scale) of the five-wafer stack, in the	
	$\hat{x},\hat{y}$ plane at the experiment zone ($\mathcal{E}$). Each wafer had
	a channel cut through it to define the electrode structure and to provide a
	path for laser beams to pass through the wafer stack. The top and bottom
	wafers provided the confining potential; the ions were trapped between these
	electrodes as indicated. The RC low-pass filters were surface-mounted to the
	filter board with gold ribbon attached by use of resistive welds. The bias
	wafer was a single electrode used to null stray electric fields along $\hat{y}$.
	Gold (represented in yellow) was coated on the top side of the trap wafers and
	wrapped around to the bottom side, and vice-versa for the bias wafer. Gold
	wire bonds connected traces on the trap wafers to traces on the filter board. }
\label{fig:trapWafers}
\end{figure}

\begin{figure}
	\includegraphics[width=0.45\textwidth]{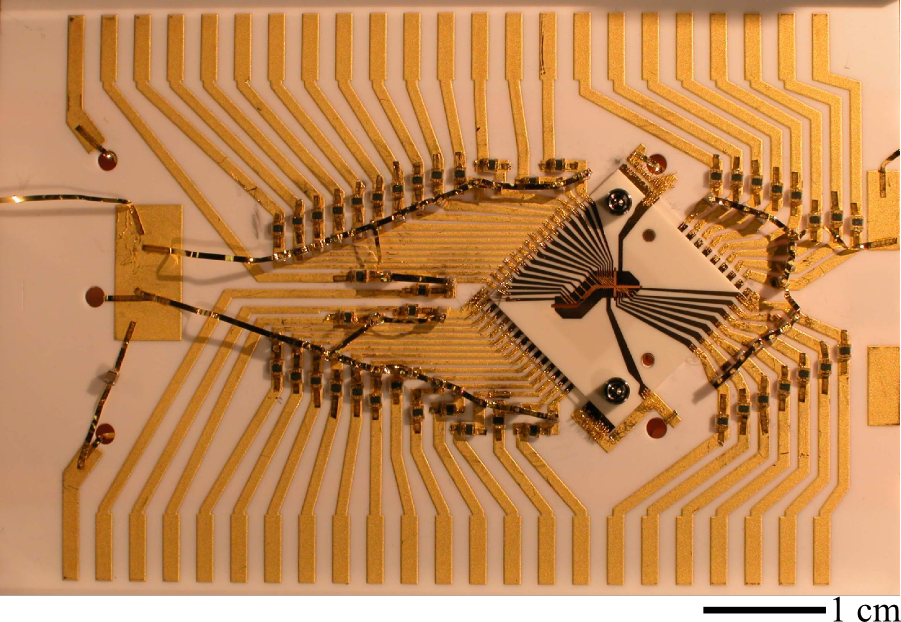}
	\caption{Top view of the filter board and trap wafers. The filter board fills
	the entire image, while the top wafer is the rotated square visible on the
	right of the image. Cap screws, visible in two corners of the top wafer, held
	the wafer stack together. Wire bonds connected the filter board traces to the top
	and bottom trap wafers. Surface-mount resistive and capacitive elements on the
	filter board provided filtering for the control potentials (see
	Fig.~\ref{fig:DCfilters}).}
\label{fig:actual_trap}
\end{figure}

The X-junction array was based on the design of previous two-layer linear RF Paul
traps~\cite{rowe2002a, barrett2004a, jost2010a}. The current trap consisted of a
stack of five high-purity alumina ($99.6\%$ Al$_2$O$_3$) wafers clamped together
(Fig.~\ref{fig:trapWafers}) with screws (visible in Fig.~\ref{fig:actual_trap}).
The trap electrodes resided in the `top' and `bottom' wafers. These wafers
($125~\mu$m thick) were laser machined to cut out `main channels' through the
wafers, with opposite sides of the channel forming rf and control electrodes.
Slits, nominally perpendicular to the main channel axes, separated the control
electrode side of the channel into a series of cantilevered structures to produce
separate control electrodes. Electrodes were formed onto the Al$_2$O$_3$ by
evaporating through a shadow-mask a $30$~nm titanium adhesion layer followed by
$0.5~\mu$m of gold, then overcoating with $3~\mu$m of electroplated gold. Care
was taken to coat all sides of each cantilevered structure to minimize exposed
dielectric that could otherwise charge and shift the potential minima in an
uncontrolled way.

A spacer wafer provided a separation of $250~\mu$m between the two trap electrode
layers. These three wafers sat atop a $500~\mu$m thick `filter board', upon which
in-vacuum RC filtering components were mounted. The `bias wafer' resembled the
`top' and `bottom' wafers but with a single continuous control electrode
extending along all sides of the main channels. The bias wafer sat below the
filter board and was used to compensate stray electric fields along $\hat{y}$.

Gauge pins were used to help align the wafers during assembly. A misalignment
error of approximately $0.22^{\circ}$ was measured between the $\hat{z}$ axes of
the two electrode wafers, and this error was included in the computer model of
the trap used to determine the appropriate transport potentials.

\begin{figure}
	\includegraphics[width=0.5\textwidth]{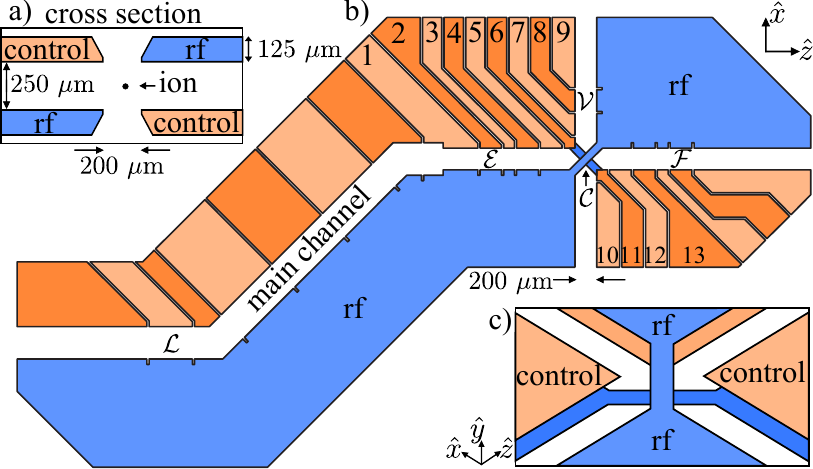}
	\caption{(a) Cross-sectional view of the two
	layers of electrodes in the X-junction array. (b) Top view of the electrode
	layout, with the rf electrodes indicated, and all other (control) electrodes
	held at rf ground. The bottom trap wafer, which sat below these electrodes,
	had a nearly identical set of electrodes but with rf and control electrodes
	exchanged across the main channel. Ions were trapped in the main channels
	between the rf and control electrodes. Forty-six control electrodes (some of
	which are numbered for reference) supported 18 different trapping zones. The
	load zone ($\mathcal{L}$), the main experiment zone ($\mathcal{E}$), the
	vertical zone ($\mathcal{V}$), the horizontal zone ($\mathcal{F}$) and the
	center of the junction ($\mathcal{C}$) are labeled. (c) Schematic of the rf
	bridges from an oblique angle (not to scale).}
\label{fig:Xtrap}
\end{figure}

The electrode layout of the array is depicted in Fig.~\ref{fig:Xtrap} and
consisted of 46~control electrodes that produced 18~possible trapping zones. The
experiment zone, $\mathcal{E}$, was chosen as the zone where the ions interacted
with lasers for cooling and qubit operations. In addition to $\mathcal{E}$, zones
$\mathcal{F}$, $\mathcal{V}$, and $\mathcal{C}$ (at the center of the junction)
composed the four destinations of the transport protocols. The final zone of
interest was the load zone, $\mathcal{L}$, where the ions were initially trapped.

The trap dimensions were similar to those in Refs.~\cite{barrett2004a,
jost2010a}. The width of the channel between the rf and control electrodes was
$200~\mu$m, except near $\mathcal{L}$, where it increased to $300~\mu$m to
increase the volume of the loading zone and, with it, the loading probability.
Most control electrodes extended $200~\mu$m along the trap axis, but those
nearest to the junction were $100~\mu$m wide to ensure sufficient control when
ions were transported in this region.

At $\mathcal{C}$, two main channels crossed to form an X-junction, and two rf
bridges connected the rf electrodes on opposite sides of that junction (one on
the top trap wafer and one on the bottom). Without such bridges, the array would
not have provided harmonic three-dimensional confinement at the center of the
junction~\cite{chiaverini2005a, wesenberg2008a}. The widths of the bridges were
$70~\mu$m, though the trapping potential was not strongly dependent on this
dimension.

\begin{figure}
\includegraphics[width=0.45\textwidth]{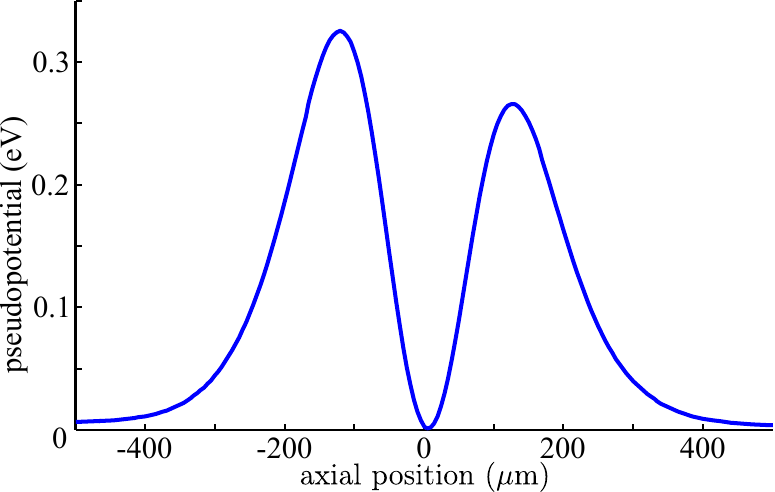}
\caption{Simulated pseudopotential barriers
along the $z$ axis produced by the rf bridges in the X-junction, with 0 being the
junction's center. Here, we assumed $V_{\mathrm{rf}} \approx
200$~V and $\OmegaRF \approx 2 \pi \times 83$~MHz. The
asymmetry between the two barriers was due to a slight misalignment of the trap
wafers.}
\label{fig:pseudopotX}
\end{figure}

These bridges introduced four axial pseudopotential barriers, one in each of the
entrances to the junction (along $\pm \hat{x}$ and $\pm \hat{z}$).
Figure~\ref{fig:pseudopotX} shows the two simulated pseudopotential barriers
along the $\hat{z}$ legs in the X-junction array going toward $\mathcal{E}$ and
$\mathcal{F}$ (the asymmetry was due to the trap misalignment mentioned above).
The height of these barriers was a significant fraction of the transverse
pseudopotential trapping depth and was approximately 0.3~eV for \Be, with rf
potential of $V_{\mathrm{rf}} \approx 200$~V (peak amplitude) and frequency
$\OmegaRF \approx 2 \pi \times 83$~MHz. At the apex of the barriers, just outside
the center of the junction, the pseudopotential was anti-confining in the axial
direction but still harmonically confining in the two radial directions. It was
possible to use the control electrodes to overwhelm this anti-confinement and
produce a 3D harmonic confining potential at all points along the axis of the
array.

Zone~$\mathcal{E}$ was positioned far ($880~\mu$m) from the junction to reduce
the residual slope of the pseudopotential barrier in this zone. The amplitude of
the axial pseudopotential at $\mathcal{E}$ was estimated, by use of computer
models, to be $2.9 \times 10^{-5}$~eV with a $8.7 \times 10^{-8}$~eV$/\mu$m axial
gradient, which would give rise to an axial 'micromotion' amplitude of 47~nm at
the drive amplitude specified above.

\begin{figure}
\includegraphics{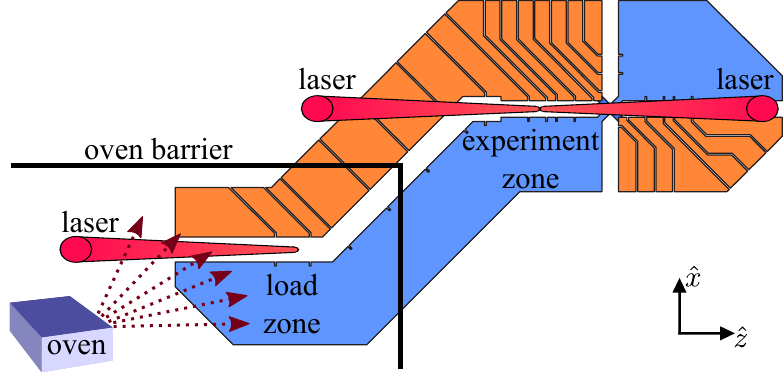}
\caption{A Be oven was positioned out of the plane of this figure in the positive $y$
direction (above the trap) and could be heated to produce a flux of neutral Be.
This Be would then travel down onto the trap, with a portion of the flux passing
into trap's main channel at the load zone. Copropagating photoionization and
Doppler-cooling laser beams intersected the Be in the load zone at 45$^{\circ}$
to the $xz$ plane of the page and parallel to $-\hat{y}+\hat{z}$. An `L'-shaped
oven barrier obscured the line-of-sight between the oven flux and the zones used
during the transport experiments to prevent neutral Be from accumulating on the
surfaces of the electrodes near the junction. This barrier was positioned just
above the trap electrodes, extending 1.6~cm along $\hat{y}$ out of the plane of
the page. Additional laser access was available for beams passing through
$\mathcal{E}$ (at 45$^{\circ}$ to the $xz$ plane) allowing for cooling,
detection, and gate operations at $\mathcal{E}$.}
\label{fig:gooseneck}
\end{figure}

Ions were loaded into the array from a flux of neutral Be that passed through
$\mathcal{L}$ and was photoionized with a mode-locked laser that after two stages
of doubling produced 235~nm resonant with the S-to-P transition of neutral Be. To
help prevent buildup of neutral Be from the beam in other regions of the array,
$\mathcal{L}$ was located sufficiently far from $\mathcal{E}$. In addition
$\mathcal{L}$ was displaced along $x$ from $\mathcal{E}$, by use of two
$135^{\circ}$ bends in the main channel, to allow an `L'-shaped stainless-steel
shield to be placed 0.5~mm above the trap wafers, preventing neutral Be from
striking the experiment zone while allowing laser access, as shown in
Fig.~\ref{fig:gooseneck}. Transporting ions through such $135^{\circ}$ bends is
relatively straightforward, and we were able to easily transport ions between
$\mathcal{L}$ and $\mathcal{E}$.

Whenever an ion was lost, a new ion was loaded into $\mathcal{L}$ and immediately
transferred to $\mathcal{E}$. It was also possible to use zone~$\mathcal{L}$ as a
reservoir zone, where extra ions were loaded and held in reserve until needed to
replace ions lost in the experiment region. This allowed the loading process to
be performed less often, which avoided heating the neutral Be oven and the
concomitant degradation of the vacuum. Potentially many ions could be
simultaneously stored in such a zone, though we only stored a small number and
did not regularly make use of this feature of the trap. By enabling a better
vacuum, a reservoir can significantly increase ion lifetime. In this scenario, it
is desirable to maintain Doppler cooling in $\mathcal{L}$ to extend the ion
lifetime.

\section{Transport Potentials}\label{sec:Waveforms}

The first demonstrations of ion transport in a multi-zone trap involved moving an
ion along a linear array~\cite{rowe2002a}. A protocol where two ions were placed
in a single trapping well and separated into two wells or combined from two wells
into a single well was also demonstrated~\cite{rowe2002a, barrett2004a}. Since
then, transport through linear arrays has been extended to other
contexts~\cite{splatt2009a, huber2008a, eble2009a}, including transport through a
junctions~\cite{hensinger2006a, blakestad2009a, amini2010a, moehring2011a} and
switching of ion order~\cite{hensinger2006a, splatt2009a}.

Here we outline the process used to calculate the time-series of control
potentials, or `waveforms', used to transport ions through the X-junction.  This
same basic procedure would be generally applicable to many ion-transport
situations. The goal was to move ions quickly, over long distances, while
maintaining low excitation of the ion's secular motion in its local potential,
and traversing non-trivial potential landscapes such as those near junctions.
Ideally, the ion should move along the axial direction of the array while
remaining at the transverse pseudopotential minimum. The control electrodes were
used to create an overall harmonic trapping well whose minimum moved along this
desired trajectory. The procedure for determining waveforms can be broken down
into four steps: modeling the trap, determining the constraints, solving for the
appropriate potentials, and assigning the time dependence of potentials.

An electrostatic model of the trap was constructed by use of boundary element
method (BEM) software~\cite{sadiku2009a, singer2010a}. For each of $N$ electrodes, the
model was run once, applying 1~V to the $n^{\textrm{th}}$ given electrode while
grounding all other electrodes. The potential resulting from each of these
voltage configurations, $\tilde{\phi}_n(\mathbf{r})$, was extracted (in the form
of a $5~\mu$m grid) in the region through which the ion would pass. These
individual electrode potentials could then be weighted by the actual voltage
applied to the electrode, $V_n$, and summed to find the total potential:
\begin{equation}\label{totalPotArray}
\Phi(\mathbf{r}) = \phi_{\mathrm{ps}}(\mathbf{r}) + \displaystyle\sum_{n=1}^N V_n
\tilde{\phi}_{n}(\mathbf{r}) .
\end{equation}
Here we have included the contribution of the rf pseudopotential
$\phi_{\mathrm{ps}}$. The pseudopotential was found by first modeling the
rf potential $\tilde{\phi}_{\mathrm{rf}}$ as if it were a \textit{static} potential
at 1~V. Then an additional step was used to convert the rf potential into a
time-independent pseudopotential by use of
\begin{equation}\label{pseudopotArray}
\phi_{\mathrm{ps}}(\mathbf{r}) = \frac{q}{4 m \OmegaRF^2}
\left( V_{\mathrm{rf}} \nabla \tilde{\phi}_{\mathrm{rf}}(\mathbf{r}) \right)^2 
,
\end{equation}
where $V_{\mathrm{rf}}$ was the peak voltage applied to the rf electrode, and
$q$ and $m$ are the charge and mass of \Be, respectively. (Throughout this
section, all $\phi$ potentials (including $\phi_{\mathrm{ps}}$) are reported as
electric potentials (in units of V) and not energy potentials (units of eV);
these are related by a factor of $q$.)

The waveform was built up from a string of individual solutions, where each
solution satisfied a set of constraints on the trapping potential centered at a
certain position. These constraints are defined below, but relate to defining the
secular frequencies and orientation of the principle axes of the potential. By
advancing that position by $5~\mu$m along the intended ion trajectory for each
subsequent solution, the series of potential steps was created that moved the
potential well along the sequence of positions. In theory, the constraints can be
set to completely define a harmonic potential localized at the desired position,
while also constraining the three secular frequencies and the orientation of the
principal axes. This would imply nine constraints, which we assume for now,
though below we will relax some of these constraints when solving for the
experiment waveforms.

To produce a trapping potential, $\Phi(\mathbf{r})$, with a minimum at
$\mathbf{r}_0=(x_0,y_0,z_0)$, we enforce
\begin{equation}\label{positionConstr}
\nabla \Phi(\mathbf{r}_0) \doteq 0,
\end{equation}
where $\doteq$ is used to mean `constrained to be true'.

The Hessian matrix,
\begin{equation}\label{Hessian2}
\mathcal{H}(\mathbf{r}_0) \equiv q \left[ {\begin{array}{ccc}
 \frac{\partial^2}{\partial x^2} & \frac{\partial^2}{\partial x \partial y} &
 \frac{\partial^2}{\partial x \partial z} \\
 \frac{\partial^2}{\partial y \partial x} & \frac{\partial^2}{\partial y^2} &
 \frac{\partial^2}{\partial y \partial z} \\
 \frac{\partial^2}{\partial z \partial x} & \frac{\partial^2}{\partial z
 \partial y} & \frac{\partial^2}{\partial z^2} \end{array} } \right]
 \Phi(\mathbf{r}_0),
\end{equation}
can be used to extract the remaining six parameters of the harmonic
potential: the eigenvalues $\lambda_i$ of $\mathcal{H}(\mathbf{r}_0)$ are related
to the secular frequencies, $\lambda_i = m \omega_i^2$ and the eigenvectors point
along the principal axes. By completely constraining the Hessian, we constrain
these quantities. Note that the Hessian is symmetric
($\mathcal{H}=\mathcal{H}^{\mathrm{T}}$), and has only six independent
entries.

It is most convenient to evaluate the Hessian in the frame of the desired
principal axes, $(x', y', z')$, in which case the Hessian constraint
equation simplifies to 
\begin{equation}\label{HessianConst}
\mathcal{H}(\mathbf{r}_0) \doteq m \left[ {\begin{array}{ccc}
 \omega_{x'}^2 & 0 & 0 \\
 0 & \omega_{y'}^2 & 0 \\
 0 & 0 & \omega_{z'}^2 \end{array} } \right]
 ,
\end{equation}
where diagonal entries constrain the desired secular frequencies $(\omega_{x'},
\omega_{y'}, \omega_{z'})$, and the off-diagonal entries constrain the principal
axes to point along  $(x', y', z')$. If the Hessian is evaluated in a different
basis, the right-hand side of Eq.~(\ref{HessianConst}) will not be diagonal, and
the frequency and axis constraints are mixed. Nonetheless, an appropriate choice for
the right-hand side can still be made in that case~\cite{brad_thesis}. From
here on, we assume $(x, y, z)=(x', y', z')$.

In the interest of compact nomenclature, it is convenient to define several
column vectors:
\begin{equation}\label{voltageVector}
\mathbf{V} \equiv \left[ {\begin{array}{ccccc}
 1 & V_{1} & V_{2} & \dots & V_{N} 
\end{array} } \right]^{\mathrm{T}}
\end{equation}
and
\begin{equation}\label{potVector}
\Psi(\mathbf{r}) \equiv \left[ {\begin{array}{ccccc}
 \phi_{\mathrm{ps}}(\mathbf{r}) & \tilde{\phi}_{1}(\mathbf{r}) &
 \tilde{\phi}_{2}(\mathbf{r}) & \dots & \tilde{\phi}_{N}(\mathbf{r})
 \end{array} } \right]^{\mathrm{T}},
\end{equation}
where $\mathbf{A}^{\mathrm{T}}$ denotes the transpose of
$\mathbf{A}$ and $\Phi(\mathbf{r}_0) = \Psi^{\mathrm{T}}(\mathbf{r}_0)
\mathbf{V}$. Finally, we define the 12-component operator
\begin{equation}\label{derivVector}
\begin{split}
\mathcal{P} \equiv & 
\bigg[ \begin{array}{cccccc}
 \frac{\partial}{\partial x} & \frac{\partial}{\partial y} &
 \frac{\partial}{\partial z} 
 & \frac{\partial^2}{\partial x^2} &  \frac{\partial^2}{\partial x
 \partial y} & \frac{\partial^2}{\partial x \partial z}
 \end{array}
\\ & \quad
\begin{array}{cccccc}
 \frac{\partial^2}{\partial y \partial x} &
 \frac{\partial^2}{\partial y^2} & \frac{\partial^2}{\partial y
 \partial z} & \frac{\partial^2}{\partial z \partial x} &
 \frac{\partial^2}{\partial z \partial y} & \frac{\partial^2}{\partial
 z^2} \end{array} \bigg]^{\mathrm{T}},
\end{split}
\end{equation}where the first three components are the gradient and the next
nine components are the Hessian operator.

The nine position, frequency, and axis constraints defined by
Eqs.~\ref{positionConstr} and~\ref{HessianConst} can be assembled into one
equation:
\begin{equation}\label{matrixConst}
\mathbf{C}_1 \bigl( \mathcal{P} \otimes
\Psi^{\mathrm{T}}(\mathbf{r}_0)\bigr) \mathbf{V}
\doteq \mathbf{C}_2,
\end{equation}
where $\mathbf{C}_1$ is a $j \times 12$ matrix and $\mathbf{C}_2$ is a $j
\times 1$ column vector, where $j=9$ for this example.

The position constraints in Eq.~\ref{positionConstr} can be reconstructed by
using $\mathbf{C}_1$ to pick out the three gradient components of $\mathcal{P}$
and $\mathbf{C}_2$ to set them to zero. The constraints in
Eq.~\ref{HessianConst} can be treated in a similar manner. Thus, to encode the
nine desired constraints, we use 
\begin{equation}\label{positionConstr2a}
\mathbf{C}_{1} = \left[ {\begin{array}{ccc@{\hspace{11pt}}ccccccccc}
 1 & 0 & 0 & 0 & 0 & 0 & 0 & 0 & 0 & 0 & 0 & 0\\ 
 0 & 1 & 0 & 0 & 0 & 0 & 0 & 0 & 0 & 0 & 0 & 0\\
 0 & 0 & 1 & 0 & 0 & 0 & 0 & 0 & 0 & 0 & 0 & 0\\ [6pt]
 0 & 0 & 0 & 1 & 0 & 0 & 0 & 0 & 0 & 0 & 0 & 0\\ 
 0 & 0 & 0 & 0 & 0 & 0 & 0 & 1 & 0 & 0 & 0 & 0\\
 0 & 0 & 0 & 0 & 0 & 0 & 0 & 0 & 0 & 0 & 0 & 1\\ [6pt]
 0 & 0 & 0 & 0 & 1 & 0 & 0 & 0 & 0 & 0 & 0 & 0\\
 0 & 0 & 0 & 0 & 0 & 1 & 0 & 0 & 0 & 0 & 0 & 0\\
 0 & 0 & 0 & 0 & 0 & 0 & 0 & 0 & 1 & 0 & 0 & 0
\end{array} } \right]
\end{equation}
and 
\begin{equation}\label{positionConstr2b}
\mathbf{C}_{2} = \left[ {\begin{array}{c}
 0 \\
 0 \\
 0 \\ [6pt]
 (m/q) \omega_x^2 \\
 (m/q) \omega_y^2 \\
 (m/q) \omega_z^2 \\ [6pt]
 0 \\
 0 \\
 0
\end{array} } \right].
\end{equation}
Additional white space has been inserted in both equations to aid the reader by
separating the position, frequency, and principal axis constraints into groups in
the vertical direction, as well as separating the gradient and Hessian
components of Eq.~\ref{positionConstr2a} in the horizontal direction.

Once $\mathbf{C}_1$ and $\mathbf{C}_2$ are determined, Eq.~(\ref{matrixConst})
can be solved for $\mathbf{V}$ by inverting $\mathbf{C}_1 \bigl( \mathcal{P}
\otimes \Psi^{\mathrm{T}}(\mathbf{r}_0)\bigr)$, thus determining the control
voltages that create the desired trapping potential. This inversion may not be
strictly possible, as is the case when the number of constraints does not equal
the number of control potentials, leading to an over- or under-determined
problem. Also, we are interested only in solutions where the magnitudes of all
control voltages are smaller than a maximal voltage $V_{\mathrm{max}}$ (for our
apparatus, $V_{\mathrm{max}} = 10$~V). To achieve this, we use a constrained
least-squares optimization algorithm, as described in Ref.~\cite{gill1981a}, to
calculate
\begin{equation}\label{LLSeqn}
\min_{\lvert V_i \rvert \le V_{\mathrm{max}}} 
\abs{ \mathbf{C}_1 \bigl( \mathcal{P} \otimes
\Psi^{\mathrm{T}}(\mathbf{r}_0)\bigr) \mathbf{V} - \mathbf{C}_2 }^2.
\end{equation}
In cases where Eq.~(\ref{matrixConst}) is over-constrained, this method yields
a ``best-fit'' $\mathbf{V}$. When Eq.~(\ref{matrixConst}) is under-constrained,
as is usually the case for large trap arrays with many electrodes, it returns a
null space in addition to $\mathbf{V}$, which can be added to $\mathbf{V}$ to
find multiple independent solutions.

Nine constraints were used above, but many are unnecessary. For QIP in a linear
trap array, constraining the axial mode frequency and orientation is often
sufficient. Parameters for the other two modes are less important and often
achieve reasonable values without being constrained, in which case they can
be omitted from the constraint matrices.

In addition to explicitly defined user constraints, there are implicit physical
and geometric constraints that must be considered. As an example, take the three
secular frequencies of the ion, $\omega_x$, $\omega_y$, and $\omega_z$. These
frequencies result from a hybrid potential that includes both pseudopotential and
control potentials. The contributions from both potentials can be separated
mathematically into components, $\omega_{\textrm{rf},i}$ and $\tilde{\omega}_i$
respectively, which add in quadrature to give the overall frequency: $\omega_i^2
= \tilde{\omega_i}^2 + \omega_{\textrm{rf},i}^2$. (An imaginary frequency
component would imply antitrapping, while a real component yields trapping.)
The control electrodes produce a quasi-static electric field, which
Laplace's equation requires to be divergenceless. This places a physical
constraint on the frequencies components due to the control potential, namely
$\displaystyle\sum_{i=1}^{3} \tilde{\omega}_i^2 = 0$. Thus, Laplace's equation
permits only certain combinations of the secular frequencies. For a
\textit{linear} Paul trap, where $\omega_{\textrm{rf},z} = 0$, the secular
frequencies must obey
\begin{equation}\label{secFreqConst}
\omega_{x}^2 + \omega_{y}^2 + \omega_{z}^2 = 2 \omega_{\mathrm{rf}}^2,
\end{equation}
where $\omega_{\mathrm{rf}}$ is the pseudopotential radial trapping frequency.

The trap geometry can place constraints on the trapping potentials, as well. For
example, in traps where the geometry contains some symmetry, the potentials must
preserve that symmetry. Care must be exercised to ensure that user-defined
constraints do not contradict physical or geometry constraints, as this will
invalidate the solution.

Though we invoke only position, frequency, and orientation constraints here,
other varieties of user-defined constraints can be easily included with this
framework, and a more complete discussion of these constraints is presented
in~\cite{brad_thesis}. The constraints used to construct the waveforms in the
X-junction array were as follows:
\begin{enumerate}
	\item The position of the potential minimum was constrained in three
	directions to be at $\mathbf{r}_0$.
	\item One of the principal axes was constrained to lie along the trap axis
	(which involves two constraints on axes orientation).
	\item The ion axial frequency was constrained (usually to 3.6~MHz).
	\item The voltages were constrained to be between $\pm 10~$V (to conform to
	the limits of the voltage supplies used in the experiment).
\end{enumerate} 
This relatively sparse set of constraints tended to give good solutions at most
locations considered. Item~4 is an inequality constraint that is easily
implemented by use of the constrained least-squares method.

When solving waveforms that transport across multiple zones, $\mathbf{V}$ can
become discontinuous from step to step, especially when transitioning between
sets of control electrodes. These jumps occur when an under-constrained problem
(with null space rank $> 0$) has multiple linearly-independent solutions and the
algorithm returns a different solution from one step to the next: during
transport there will be some position at which it is suddenly easier to produce
the desired potential using a new combination of electrodes. In principle, such
jumps should not have adverse effects on the potential at the ion, as the
potentials on both side of the jump fulfill the same constraints and should
transition smoothly. However, since the potentials on the electrodes are
filtered, we would expect the potential at the ion to experience a transitory
jump during the transition.

These solution jumps can be handled by various means. We used the constrained
least-squares method to seed each new solution with the solution of the previous
step while introducing a cost for deviating from the previous solution by
replacing Eq.~\ref{LLSeqn} with
\begin{equation}\label{LLSeqn2}
\min_{\lvert V_i \rvert \le V_{\mathrm{max}},\ \lvert V_i -
V_{i,\mathrm{last}}\rvert \le \alpha} 
\abs{ \mathbf{C}_1 \bigl( \mathcal{P} \otimes
\Psi^{\mathrm{T}}(\mathbf{r}_0)\bigr) \mathbf{V} - \mathbf{C}_2 }^2 ,
\end{equation}
for a positive constant $\alpha$. This removes the need for iteratively
choosing weights to keep the voltages within bounds, as suggested in
Ref.~\cite{singer2010a}. This forced the jump transition to be extended over
multiple steps, rather than allowing a discontinuous jump. Another approach is to
average the two $\mathbf{V}$'s on each side of the discontinuity, taking
advantage of the linearity of the equations, to produce an intermediate solution
that still satisfies the constraints~\cite{uys2009a}. Performing several steps of
such averaging will smooth the jump. Alternately, trial and error can often be
used to determine a set of constraints that does not produce a jump, but this can
require significant effort.

\subsubsection{Transport timing}

If the spatial interval between steps in the waveform is small enough, the
potential, once applied to the electrodes, will move smoothly from step to
step~\cite{brad_thesis}. The velocity of the potential well (and, thus, the ion)
is controlled by the rate at which the waveform steps are updated on the
electrodes. In our case, the control potentials were supplied by
digital-to-analog converters (DACs) that had a constant update rate $\OmegaDAC =
480$~kHz and the number of update steps was adjusted to change the velocity.

Different velocity profiles have been considered for minimizing excitation while
transporting~\cite{reichle2006b, hucul2008a}. In this report, the ions were
usually transported by use of a constant velocity with equally spaced waveform
steps. This could potentially lead to the ion being `kicked' as the velocity
jumps at the beginning and end of the transport, resulting in motional
excitation. However, these velocity jumps were smoothed by low-pass filters
placed on the control potentials (see Sec.~\ref{subsec:DACNoise}). A smoother
`sinusoidal' velocity profile was also tested but was abandoned after observing
no discernible difference in the amount of motional excitation by use of the
different profiles. This suggests that both transport protocols were well within
the adiabatic regime at the speeds used.

Low-pass filtering (160~kHz corner in our case) can also potentially distort the
waveforms when transporting quickly, placing an upper limit on the ion speed.
However, the practical speed limit was set by the combination of the
maximum update rate of the digital-to-analog converters and the number of update
points required to accurately produce a continuous harmonic potential in the
region of the pseudopotential barrier. This limit was experimentally determined
for each waveform by adjusting the number of update points until minimum motional
excitation was achieved. If faster DACs are available and distortion of the
waveforms due to low-pass filtering is of concern, the waveform can be
pre-compensated to account for these distortions and produce the desired waveform
at the ion.

\begin{figure}
\includegraphics[width=0.5\textwidth]{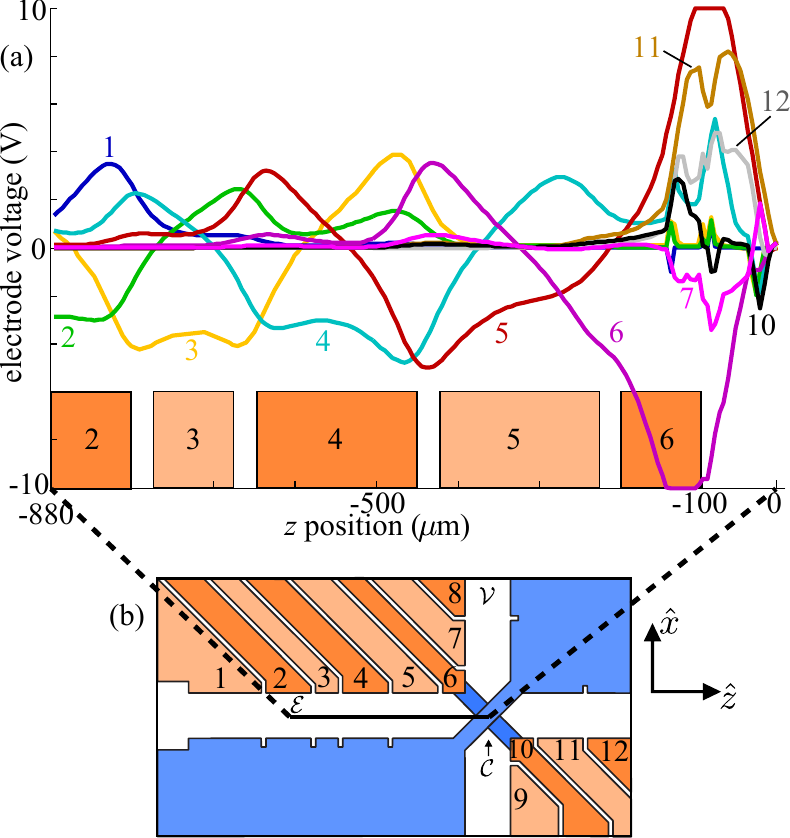}
\caption{(a) The waveform (as a function of position rather than time) used when
transporting an ion from the experiment zone, located at $z=-880~\mu$m, to the
center of the junction at $z=0~\mu$m. We plot voltage versus the $z$ position of
the minimum of the trapping potential during the transport. 
The locations of the electrodes near the
junction are depicted, along with their electrode number, by the rectangles in
the bottom of the figure. The region from $- 100$ to $0~\mu$m is inside of the
junction. The voltage traces are numbered to show which electrode they correspond
to. Electrodes 8 and 9 remain near 0~V and are omitted for clarity. In
addition, the potentials applied to the control electrodes on the bottom wafer
are not displayed, as they are nearly identical to those applied on the top
wafer. (b)~A schematic of the trap, showing the range over which this waveform
transported.}
\label{fig:wvf}
\end{figure}

The waveforms used to transport from $\mathcal{E}$ to $\mathcal{C}$ are displayed
in Fig.~\ref{fig:wvf} as a function of the position of the minimum of the
trapping potential (the ion's location). The potentials applied to the lower
trap-wafer control electrodes (on opposite sides of the main channel) were nearly
identical and are omitted for clarity. These waveforms could be run left to right
to transport an ion $880~\mu$m from $\mathcal{E}$ to $\mathcal{C}$, or they could
be run in reverse. The waveforms that transported ions into the other two
branches of the junction (to $\mathcal{F}$ and $\mathcal{V}$) were similar to
this waveform due to the approximate symmetry of the trap.

In a typical transport, the potential minimum was moved at a constant velocity,
and there was a direct linear relationship between the location of the minimum
(horizontal axis of Fig.~\ref{fig:wvf}(a)) and the time elapsed since the
beginning of the transport. The typical transport duration for the waveforms in
Fig.~\ref{fig:wvf}(a) was approximately 165~$\mu$s, with 50~$\mu$s to cross the
pseudopotential barrier.

Some control potentials reached the $\pm 10$~V limit placed by use of the
constrained least-squares method while traversing the pseudopotential barrier
near the junction. Other control potentials had sharp and abrupt changes, which
resulted from the constraint in Eq.~\ref{LLSeqn2} that prevented `solution
jumping' by defining how much a given waveform step can deviate from the previous
step. Instead of jumping, the voltages ramped linearly over several steps.
Although these individual potentials were not smooth in time, they were
continuous, which was sufficient to ensure that the overall potential experienced
by the ion evolved smoothly.

\begin{figure}
	\includegraphics[width=0.45\textwidth]{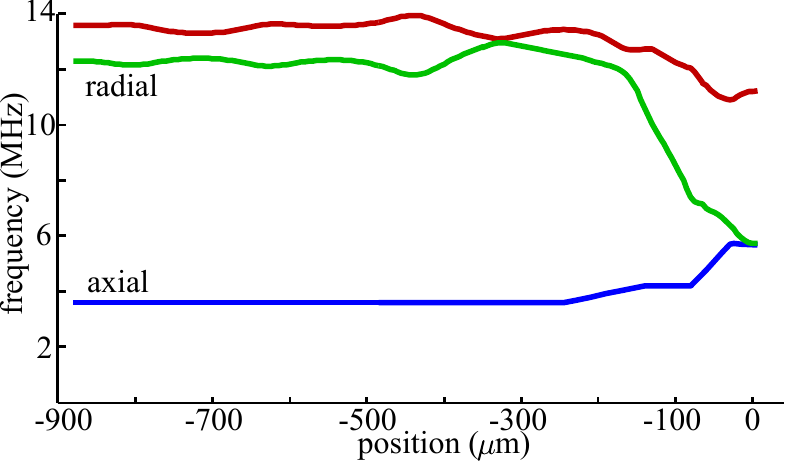}
	\caption{Predicted secular frequencies as a function of position
	corresponding to the waveform in Fig.~\ref{fig:wvf}. The axial frequency along
	$\hat{z}$ was constrained to be 3.6~MHz during the majority of the transport,
	while the radial frequencies were unconstrained. As the ion ascended the
	pseudopotential barrier, the axial frequency linearly ramped up to 4.2~MHz.
	Beyond the apex of the barrier, a second linear ramp was applied to bring the
	frequency up to 5.7~MHz. As the ion approached the center of the junction, the
	$x$ and $z$ frequencies became nearly degenerate.}
\label{fig:wvfFreq}
\end{figure}

The axial frequency was chosen to be 3.6~MHz and was held constant during much of
the transport starting at $\mathcal{E}$ and moving toward $\mathcal{C}$
(Fig.~\ref{fig:wvfFreq}). The frequency was (adiabatically) linearly ramped to
4.2~MHz as the ion approached the apex of the pseudopotential barrier, making the
ion less susceptible to rf-noise heating of the secular motion (see
Sec.~\ref{subsec:rfNoise}). The value 4.2~MHz was the maximum axial frequency
attainable at the apex due to the strong anti-confinement of the pseudopotential
at that location and the $\pm 10$~V limit of the DACs providing the
control potentials. The axial frequency then continued to increase as the ion
descended the barrier, reaching a final value of 5.7~MHz at $\mathcal{C}$. At
this location, all control potentials were 0~V and the pseudopotential provided
all the trapping, resulting in near-degenerate 5.7~MHz confinement along the
$\hat{x}$ and $\hat{z}$ directions, while the $\hat{y}$ secular frequency was
11.3~MHz. When transporting multiple ions in the same potential well, it would be
preferable to break the frequency degeneracy at $\mathcal{C}$ to ensure
well-defined axes for the ions. In practice, the motional excitation rates
when moving pairs of ions were still relatively low despite the near degeneracy
at $\mathcal{C}$ (see Table~\ref{tab:heatingrates2ions}).

\section{Transport Experiments}\label{sec:Experiment}

The transport experiments were performed with \Be ions inside a vacuum system
with a pressure of $p < 5 \times 10^{-11}$~Torr~$=7\times 10^{-9}$~Pa. A $1.3
\times 10^{-3}$~T magnetic field was applied to split the Zeeman states, and the
ions were optically pumped to the $^2 S_{1/2} \left| F=2, m_F=-2 \right>$ state
(henceforth $\ket{2,-2}$).   Manipulation of the \Be ion motional and internal
states used the techniques of Refs.~\cite{monroe1995a,bible}. Two-photon
stimulated-Raman transitions enabled coherent transitions between the qubit
states $\ket{2,-2}$ and $\ket{1,-1}$ at frequency $\omega_0 \approx 2 \pi \times
1.28$~GHz. In addition, by tuning the difference frequency of the Raman beams to
$\omega_0 \pm \omega_z$ , it was possible to drive a blue(red)-sideband
transition: $\ket{2,-2}\ket{n} \leftrightarrow \ket{1,-1}\ket{n \pm 1}$. Here
$\ket{n}$ is a Fock state of a selected motional mode. Ground-state cooling was
performed by use of a series of red-sideband pulses, followed by repeated optical
pumping to $\ket{2,-2}$. State detection was performed using state-dependent
resonance fluorescence, where predominantly the $\ket{2,-2}$ state fluoresces.

Each transport began by cooling an ion (or ion pair) in \tE\ to the motional
ground state. The ion was then transported into or through the junction and
returned to \tE. Three transports patterns were used: \tECE\ moved to
$\mathcal{C}$ and back, while \tECFCE\ and \tECVCE\ moved to $\mathcal{F}$ and
$\mathcal{V}$, respectively, before returning to $\mathcal{E}$. The \tECE\
transport moved the ion $1.76$~mm, while \tECFCE\ and \tECVCE\ moved the ion
$3.52$~mm and $2.84$~mm, respectively. Once the ion returned to $\mathcal{E}$,
the motional excitation was determined by measuring the asymmetry in red- and
blue-sideband Raman transitions~\cite{monroe1995a, turchette2000a}.

To determine the single-ion transport success rate for \tECFCE\ transports, two
sets of 10,000 consecutive transport experiments were
performed~\cite{blakestad2009a}, but with the imaging system focused on \tE\ in
the first set and on \tF\ in the second. The first set verified that the ion
successfully returned to \tE\ every time. The second set verified that the ion
always reached \tF\ at the intended time. Together, these sets of experiments
imply the success rate for going to \tF\ and returning to \tE\ exceeded 0.9999.
The procedure was repeated for \tECVCE, with the same result. The \tECE\
transport can not be verified in the same manner because the bridges obscure the
ion at \tC, but since the ion must transport through this location to reach \tF\
and \tV, the reliability should be no worse.

Ion lifetime, and thus transport success probability, was ultimately limited by
ion loss resulting from background-gas collisions~\cite{bible}. With this in
mind, the ion loss rate during transport was not larger than that for a
stationary ion ($\sim 0.5/$hr). Having observed millions of successive round
trips for all three types of transport, combining all losses implies a transport
success probability of greater than 0.999995\footnote{For the 0.999995 success
probability figure, we only verified that each transport successfully returned
the ion to \tE, not whether the ion successfully moved the ion through the
junction. Given the low rates of motional excitation, and the fact that we did
verify that the ions move through the junction using 10,000 experiments, it is
reasonable to assume the ion did travel through the junction if the ion
successfully returned to \tE.}. Since transport comprised a small fraction of the
total experiment duration, many of these losses likely occurred when the ion was
not being transported. In one instance, more than 1,500,000 consecutive \tECE\
transports were performed with a single ion.

Loss rates for transported ion pairs were again comparable to stationary pairs
($\sim 2$~per hour). Absolute pair loss rates were higher than those for single
ions, presumably due to multi-ion effects~\cite{bible, walther1993a}.

\section{Excitation of the Secular
Motion}\label{sec:HeatingSources}

Excitation of the ion's motion during transport was attributed to two main
mechanisms: one due to rf noise near $\OmegaRF$ and the other due to excitation
from the digital-to-analog converters (DACs).

\subsection{rf-noise heating}\label{subsec:rfNoise}

Consider a trapping rf electric field with an additional sideband term,
\begin{equation}\label{N1efield}
\mathbf{E}_{\textrm{rf}}(\mathbf{r},t)=\mathbf{E}_{0}(\mathbf{r}) \left[
\cos{ \OmegaRF t }+\xi_{\textrm{N}} \cos{(\OmegaRF \pm \omega_z)t} \right],
\end{equation}
where $\xi_{\textrm{N}} \ll 1$, and $\OmegaRF \pm \omega_z$ is at one of
the two axial motional sidebands of the ion. In~\cite{blakestad2009a}, it was
shown that the two terms will beat at $\omega_z$ to produce a force that can
excite the ion's motion. If the second term is not coherent, but instead is
broad-spectrum noise, this will lead to excitation of the axial motion at a
rate of
\begin{equation}\label{N1heatingrate}
\begin{split}
\dot{\bar{n}}_z = & ~ \frac{q^4}{16
m^3 \OmegaRF^4 \hbar \omega_z} \left[ \frac{\partial}{\partial z}
E^2_{0}(z) \right]^2 \times \\ & \quad \left(
\frac{S_{V_{\textrm{N}}} (\OmegaRF + \omega_z) }{V_{\mathrm{rf}}^2} +
\frac{S_{V_{\textrm{N}}} (\OmegaRF - \omega_z)}{V_{\mathrm{rf}}^2}
\right) ,
\end{split}
\end{equation}
where $S_{V_{\textrm{N}}} (\OmegaRF \pm \omega_z)$ is the voltage-noise
spectral density at either the upper or lower rf sideband, and $V_{\mathrm{rf}}$
is the amplitude of the trapping rf potential being applied to the rf electrodes.
$E_0(z)$ is the axial rf electric field amplitude at the location of the ion.
This heating mechanism is proportional to the \emph{slope} of the pseudopotential
and is significant only in places with a large slope, such as the pseudopotential
barriers  near the junction (but not in, for example,~\tE).

\begin{figure}
\includegraphics{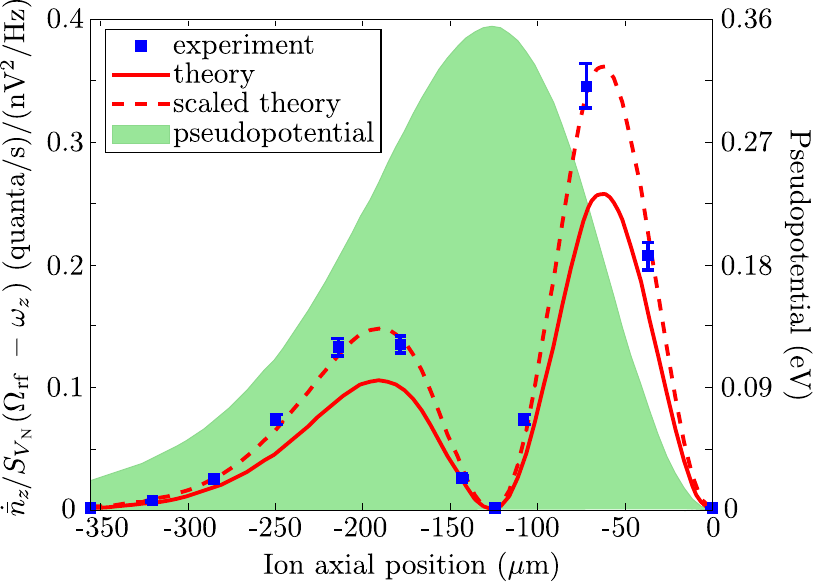}
\caption{The ratio of heating rate $\dot{\bar{n}}_z$ to voltage noise spectral
density $S_{V_{\textrm{N}}} (\OmegaRF - \omega_z )$ for various locations along
the trap axis (\tC~is located at 0~$\mu$m). This figure is reproduced from
Ref.~\cite{blakestad2009a}. The theoretical prediction used a pseudopotential
modeled from electrode geometry and is shown both with and without a scaling
parameter ($= 1.4$). The simulated pseudopotential is overlaid in the background,
in units of eV. Since heating was gradient dependent, we saw very little heating
at the peak of the pseudopotential barrier, even though this was the point of
maximum (axial) rf electric field and therefore maximum axial rf micromotion.
Nearly identical pseudopotential barriers were present on the other three legs of
the junction.}
\label{fig:rfheating}
\end{figure}

This heating mechanism was verified in Ref.~\cite{blakestad2009a} by measuring
the heating rate at various locations along the pseudopotential barrier between
\tE\ and \tC, while spectrally-dense white noise (centered on the lower sideband,
$\OmegaRF - \omega_z$) was injected onto the trap rf drive.
Figure~\ref{fig:rfheating} plots the ratio of measured heating rate to estimated
injected~$S_{V_{\textrm{N}}}$ and theoretical values of this ratio according to
Eq.~(\ref{N1heatingrate}) based on simulations of trap potentials, for the ion
held at several positions between \tE~and \tC. A plot with the theoretical values
multiplied by a scaling factor ($=1.4$) is also included.  The deviation of the
scaling factor from 1 is not unreasonable due to the difficulty of accurately
measuring a variety of experimental parameters.

The motional excitation for full junction transports to $\mathcal{C}$ was
observed to decrease as the ion speed was increased, which minimized the exposure
to the rf noise while on a pseudopotential slope. This continued up to a maximum
speed limit, due to the slow DACs, above which the other excitation mechanism
(below) began to dominate. At the optimum speed, the ion spent only approximately
$50~\mu$s on each barrier (above 10\% of the barrier height).

Another approach to mitigate rf noise is to suppress the sideband noise with
better filtering of the applied rf trapping potential. In
Ref.~\cite{blakestad2009a}, the large rf potential ($V_{\mathrm{rf}} \approx
200$~V$_{\mathrm{peak}}$ at $\OmegaRF \approx 2 \pi \times 83$~MHz) was provided
by a series of tank resonators, which suppressed noise at the motional sidebands
($\pm 3.6$~MHz). The primary resonator was a quarter-wave step-up
resonator~\cite{jefferts1995a} with a loaded $Q = 42$ and corresponding bandwidth
(FWHM) of 2~MHz. This resonator extended into the vacuum, with the trap attached
to the voltage anti-node. A second half-wave resonator with $Q = 145$ was
attached, in series, to the input of the primary resonator, with a 3~dB
attenuator in between to decouple the two resonators. The resonant frequencies of
the two resonators were tuned to be equal. This network resulted in an estimated
ambient $S_{V_{\textrm{N}}} (\OmegaRF \pm \omega_z)$ of -177~dBc at the ion. In
the work reported here, the second resonator was replaced with a pair of
half-wave tank resonating cavities. This filter pair provided more than 38~dB
suppression at frequencies $\OmegaRF \pm 2 \pi \times 3.6$~MHz (when not coupled
to the primary resonator), an additional suppression of approximately 10~dB over
the half-wave filter used in Ref.~\cite{blakestad2009a}. $S_{V_{\textrm{N}}}$ at
the ion was not re-measured with this new filter pair, but observed reductions in
excitation during transport were consistent with a 10~dB drop in rf noise.

\subsection{DAC update noise}\label{subsec:DACNoise}

Another primary source of motional excitation was attributed to the 16-bit,
$\pm 10$~V DACs that supplied the waveform potentials to the electrodes. The DAC
voltages were updated at a constant rate $\OmegaDAC$ ($\leq 500$~kHz), resulting
in Fourier components that could excite the ion's motion if $2 \pi \times
\OmegaDAC = \omega_z/J$, for any integer $J$.

This effect was observed by first preparing the ion in the motional ground state
at \tE\ and then transporting toward \tC. Instead of proceeding all the way to
\tC, the transport was stopped (at $z=-300~\mu$m) before the axial frequency
began to ramp up. Thus, the local potential-well frequency remained approximately
constant at $\omega_z = 2 \pi \times 3.6$~MHz. The ion was then returned to~\tE.
A red-sideband Raman $\pi$-pulse for $n = 0$ to $n = 1$ excitation was applied to
determine if the ion remained in the ground state~\cite{monroe1995a,
turchette2000a}. If the ion was excited out of the ground state during transport,
the Raman pulse had a certain probability to transfer the ion into the
$\ket{1,-1}$ state, which did not fluoresce during detection. If the ion remained
in the ground state, this side-band pulse had no effect and the ion remained in
the bright $\ket{2,-2}$ state. Thus, fluorescence detection after the side-band
pulse could distinguish an excited ion from a non-excited ion.

\begin{figure}
\includegraphics{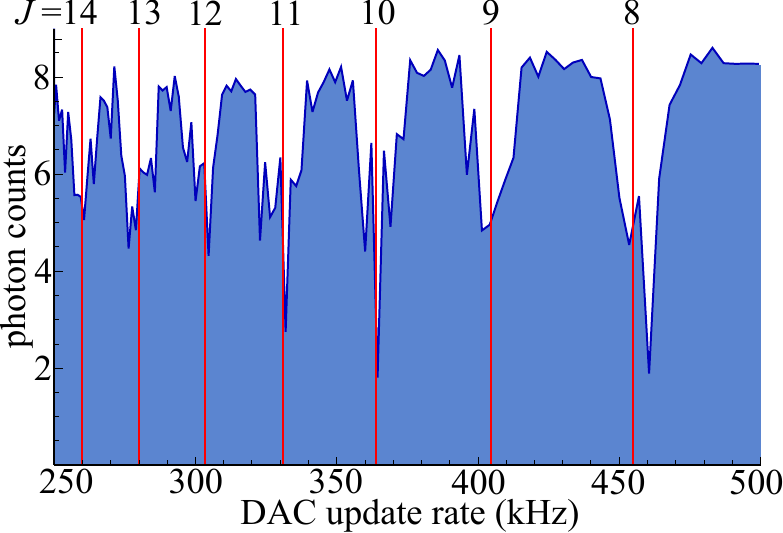}
\caption{Plot showing the the number of fluorescence photons detected in a
duration of $200~\mu$s following round-trip transport and a subsequent red
sideband pulse, for various DAC update rates $\OmegaDAC$. Before transport,
the ion was prepared in the motional ground state and then transported through a
specific waveform, where $\omega_z=2\pi \times 3.6$~MHz was maintained during
the entire transport. If the ion remained in the ground state after transport,
the ion fluorescence was at its maximum value (approximately 8~photon counts
detected), but when the ion became motionally excited, the fluorescence dropped.
As can be seen, the motion was excited at specific update frequencies that
correspond to $\OmegaDAC = \omega_z/(2 \pi J)$ for $J=8$ to $14$ (marked by the
vertical red lines).}
\label{fig:DACheating}
\end{figure}

This experiment was performed for various values of $\OmegaDAC$, and the results
are shown in Fig.~\ref{fig:DACheating}. It was difficult to extract the ion's
exact motional state after the transport, but the correlations between the ion's
motional excitation and $\omega_z$ corresponding to a harmonic of $\OmegaDAC$
were evident. The energy gain exhibited a resonance at several values for
$\OmegaDAC = \omega_z/(2 \pi J)$ with $J=8$ to $14$. When the number of update
steps was increased, while the update rate was held constant (which resulted in
an increased transport duration), the bandwidth of these resonances decreased, as
expected from a coherent excitation.

Use of an update rate that was incommensurate with the motional frequency
($\OmegaDAC \neq \omega_z/(2 \pi J)$) minimized this energy gain. However,
increasing the transport speed (using the same update rate) required a reduction
in the number of waveform steps, which caused the resonances to broaden.
Minimizing the rf-noise heating required fast transport, so at the speed that
gave the lowest rf-noise excitation rates, the DAC heating resonances were so
broad that they overlapped, and there was no achievable $\OmegaDAC$ that would
not result in energy gain. The DAC heating effect was further compounded by the fact
that the axial frequency was not constant during a full junction transport,
making it impossible to achieve $\OmegaDAC \neq \omega_z/(2 \pi J)$ for any
constant $\OmegaDAC$. The update frequency $\OmegaDAC = 480$~kHz appeared to be
most favorable and was used for the results here and in
Ref.~\cite{blakestad2009a}.

\begin{figure}
	\includegraphics[width=0.5\textwidth]{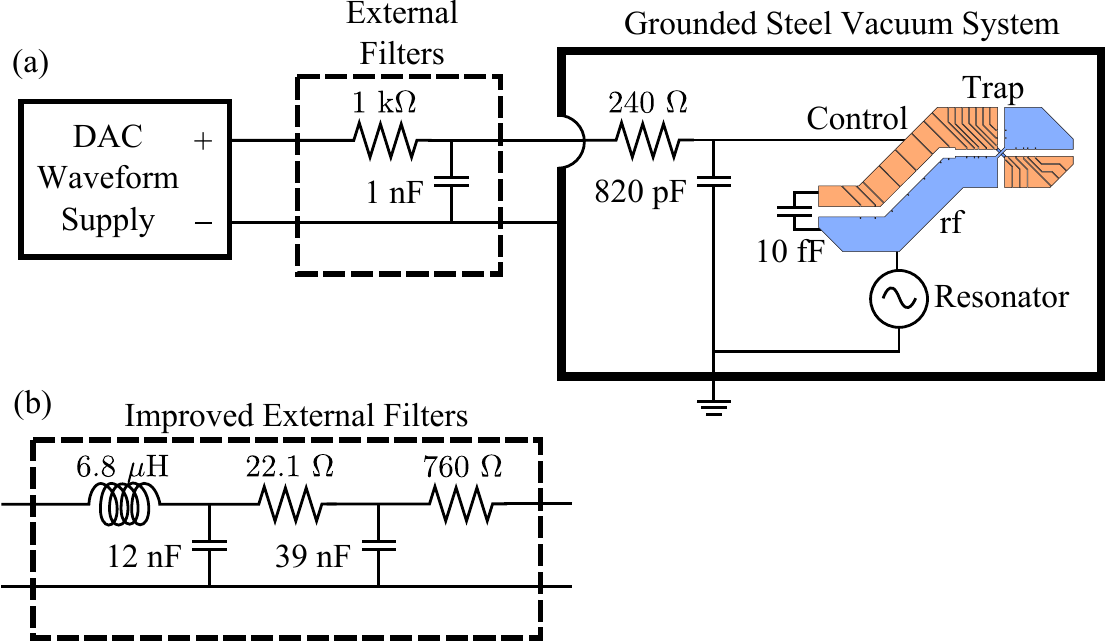}
\caption{The control
voltages were provided by 40 independent DACs (only one shown here). The DAC
output was filtered prior to being applied to the trap electrodes, through a
two stage filter seen in (a). The control potentials were referenced to the
grounded vacuum system, which served as the rf ground as well. (b)~After
DAC-update noise was observed to excite the secular motion of the ions, the
external filters were replaced by the approximate $3^{\mathrm{rd}}$-order
`Butterworth' filter shown here.}
\label{fig:DCfilters}
\end{figure}

Faster DACs capable of $\OmegaDAC>\omega_z/2 \pi$ should significantly suppress
this motional excitation. Alternatively, aggressive filtering of the DAC output
can combat this problem. The results in Ref.~\cite{blakestad2009a} used the RC
filter network shown in Fig.~\ref{fig:DCfilters}(a), which provided suppression
by two orders-of-magnitude over the range of $\omega_z/2\pi$ values used
during transport (3.6 to 5.7~MHz), but was not sufficient to completely suppress
the DAC heating. Increasing the RC time constant would increase the filtering but
would also slow down the rate at which the ion can be transported. Instead, these
simple RC filters were replaced with the approximately third-order
Butterworth filter~\cite{tietze2008a} shown in Fig.~\ref{fig:DCfilters}(b). (The
output impedance of the DAC was $< 0.1~\Omega$ and contributed minimally to the
filter response.)

\begin{figure}
	\includegraphics{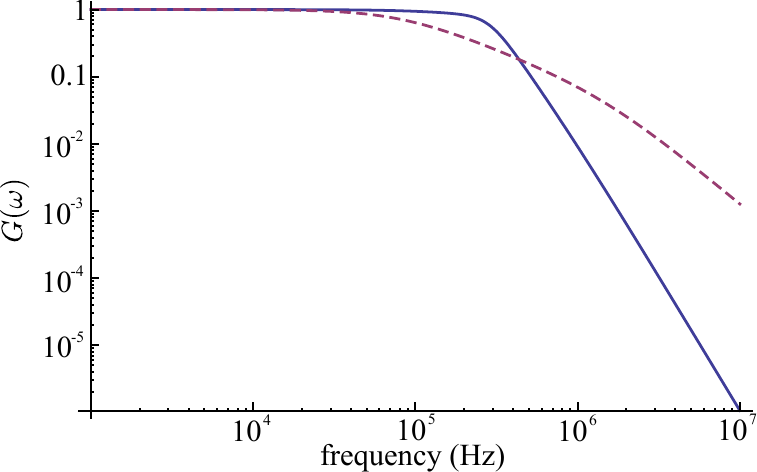}
\caption{Theoretical 
transfer function $G(\omega)$ versus frequency for the original RC filter
(dashed) and the improved approximately-`Butterworth' filter (solid), where the
new filter had a faster roll-off at high frequency. Since the pass-band
extended farther for the new filter, the transport speed could be increased
while providing more filtering at the secular frequency (3.6~MHz to 5.7~MHz).
Both traces include the RC components inside the vacuum. }
\label{fig:filterReal}
\end{figure}

A Butterworth filter has a frequency response given by
\begin{equation}\label{ButtGenGain} G(\omega) =\frac{1}{\abs{B_n(i
\omega/\omega_0)}}=\frac{1}{\sqrt{1+ (\omega/\omega_0)^{2n}}},
\end{equation}
where $B_n(s)$ is the $n^{\mathrm{th}}$-order Butterworth polynomial and
$\omega_0$ is the corner frequency. If $n=1$, the frequency response
reduces to a RC frequency response. In the experiments here, such higher-order
filters provide stronger noise suppression at $\omega_z$ while still allowing
fast transport. A comparison of the theoretical response functions for the RC
filters used in Ref.~\cite{blakestad2009a} and the Butterworth filters used here
is shown in Fig.~\ref{fig:filterReal}. The internal vacuum RC components already
on the filter board were taken into account when planning the Butterworth
filter, but the external filter components were designed to dominate the
filter's response in the frequency range of concern. Thus, the filter was
approximately third-order, despite the presence of four components
(including the filter board capacitor) with frequency-dependent impedances. The
new filters increased the noise filtering by 22~dB at 3.6~MHz and 26~dB at
5.7~MHz. Furthermore, the electric-field noise at the ion due to Johnson noise in
the resistive elements of the new filters was less than that for the previous
filters for all frequencies of interest. For the transport durations used, these
filters did not appreciably distort the waveform.

\subsection{Anomalous noise heating}\label{subsec:Anomalous}

The `anomalous heating', which is thought to arise from noisy electric potentials
on the surface of the trap~\cite{turchette2000a}, was measured to be 40~quanta/s
for $\omega_z/2 \pi = 3.6$~MHz at \tE\ and was not a significant source of
excitation during the transport experiments. For example, we estimate that it
should have contributed only 0.007~quanta for \tECE~transport. To compare the
measurements of various ion traps, it has been common to express the  heating in
terms of the electric field noise with the expression~\cite{turchette2000a}:
\begin{equation}\label{AnomHeating1} \dot{\bar{n}}_z = \frac{q^2}{4 m \hbar
\omega_z} S_E (\omega_z) ,
\end{equation}
where $S_E (\omega_z)$ is the spectral density of electric field fluctuations at
the secular frequency. From the results here we found $S_E(2 \pi \times
3.6~$MHz$) = 2.2 \times 10^{-13}~(\mathrm{V}/\mathrm{m})^2 \mathrm{Hz}^{-1}$,
where the distance of the ion to the nearest electrode surface was $160~\mu$m. 
This result, when compared to other traps as in Refs.~\cite{amini2010a,
deslauriers2004a, epstein2007b, daniilidis2011a}, was significantly below that of
most other room-temperature ion traps. The cause of this relatively low heating
rate is not known, but surface preparation could be a contributing factor.

\subsection{Other heating mechanisms}\label{subsec:otherNoise}

In the experiments here, the transport was slow and the trapping potential
changed slowly compared to the motional frequencies, so we did not expect
non-adiabatic excitation of the motion. This was supported by observations that
the excitation did not decrease as the transport was slowed.  For very slow
transport, the heating actually increased because the ion spent more time
crossing the rf barriers, resulting in increased rf-noise heating. Furthermore,
no reduction in heating was observed when a gradual (sinusoidal) velocity profile
was used instead of a constant velocity over the entire transport.

The waveforms were produced assuming a specific value of $V_{\mathrm{rf}}$ and
corresponding pseudopotential. In theory, if the actual $V_{\mathrm{rf}}$ does
not match the assumed $V_{\mathrm{rf}}$, the axial trapping potential will not be
as intended at the barriers. In practice, there was an optimal value for the rf
power which resulted in the lowest excitation rates and likely corresponded to
the assumed $V_{\mathrm{rf}}$. The rf power was prone to slow drifts over many
minutes (likely due to temperature drifts in the resonators) which resulted in
modest increases in motional excitation; it was necessary to occasionally adjust
the rf power (every 10 to 30~minutes) and hold it to within~$<1~\%$ to achieve
the lowest motional-excitation rates. In practice, this was performed by ensuring
that the radial secular frequencies at \tE\ remained constant.

\subsection{Motional excitation rates}\label{subsec:heatingRates}

The motional excitation for single-ion transports was measured by use of sideband
asymmetry measurements~\cite{monroe1995a} after a single pass through the
junction, and the results are summarized in Table~\ref{tab:heatingrates}. These
results were significantly better than those in Ref.~\cite{blakestad2009a}, which
are listed for comparison. In Ref.~\cite{blakestad2009a}, rf noise was estimated
to contribute 0.1 to 0.5~quanta of excitation per pass over a pseudopotential
barrier, which explained between 3 and $30\%$ of the excitation seen. The
remainder of the excitation was attributed primarily to DAC update noise. The new
trap rf filters and control electrode Butterworth filters produced the observed
reduction in excitation rates.

\begin{table}
\centering
\begin{tabular}{|c|c|c|c|c|}
 & \multicolumn{2}{c}{This work} &
 \multicolumn{2}{c}{Ref.~\cite{blakestad2009a}} \\
\cline{2-5}
& Duration & Energy gain & Duration & Energy gain \\
Transport \quad  & ($\mu$s) & (quanta/trip) & ($\mu$s) & (quanta/trip) \\
\hline
\tECE & 350 & $0.053 \pm 0.003$ & 310 & $3.2 \pm 1.8$ \\
\tECFCE & 910 & $0.18 \pm 0.02$ & 630 & $7.9 \pm 1.5$ \\
\tECVCE & 950 & $0.18 \pm 0.02$ & 870 & $14.5 \pm 2.0$ \\
\end{tabular}
\caption{The axial-motion excitation $\Delta \bar{n}$ for a single \Be ion
for three different transports through the X-junction. The results of this work,
as well as that of Ref.~\cite{blakestad2009a}, are given for
comparison\protect\footnote{The transport durations given in
Ref.~\cite{blakestad2009a} were reported  in error. The correct values are
$140~\mu$s for transporting from \tE\ to \tC, $300~\mu$s to go from \tE\ to \tV,
and $420~\mu$s to go from \tE\ to \tF. This error did not affect any other
results in Ref.~\cite{blakestad2009a}.}. The transport duration includes
$20~\mu$s for the ion to remain stationary at the intermediate destination
($30~\mu$s for the data from Ref.~\cite{blakestad2009a}), before returning to
$\mathcal{E}$. The energy gain per trip is stated in units of quanta in a
3.6~MHz trapping well where $\Delta \bar{n}=0.1$~quantum corresponds to
1.6~neV.}
\end{table}

The transport durations, which were optimized for minimal excitation, are also
given in Table~\ref{tab:heatingrates}. The tabulated durations correspond to the
full transport duration including returning to  \tE\ (rather than the
half-transport reported in Ref.~\cite{blakestad2009a}. The durations also include
a $20~\mu$s wait at the half-way point (\tC, \tF, or \tV, depending on the
transport) for the new results and a $30~\mu$s wait for those from
Ref.~\cite{blakestad2009a}.

Moving pairs of ions in the same trapping well would be useful for both
sympathetic cooling and efficient ion manipulation~\cite{kielpinski2002a}. This
type of transport was demonstrated by use of pairs of \Be ions and the measured
motional excitation is reported in Table~\ref{tab:heatingrates2ions}. Excitation
in both the center-of-mass (COM) and stretch modes was measured. Additional
heating mechanisms for multiple ions~\cite{bible, walther1993a} may explain the
higher energy gain observed for the pair. For \tECVCE~transport, the two-ion
crystal must rotate from the $\hat{z}$ axis to the $\hat{x}$ axis and back.  For
the waveforms used, the potential was nearly the same in the $\hat{x}$ and
$\hat{z}$ directions at~\tC. Therefore the axes were not well defined throughout
the transport, which can lead to an uncontrolled rotation of axes. It is possible
that the discrepancy in the excitation between \tECFCE\ and \tECVCE\ for two ions
may have resulted from this uncontrolled rotation at~\tC.

\begin{table}
\centering
\begin{tabular}{|c|c|c|c|}
Transport & \multicolumn{3}{c}{Energy
gain (quanta/trip)} \\
\hline
 & This work & This work & Ref.~\cite{blakestad2009a} \\
 & COM & Stretch & COM \\
\cline{2-4}
\tECE & ~$0.39 \pm 0.03$~ & ~$0.13 \pm 0.02$~ & ~$5.4 \pm 1.2$~ \\
\tECFCE & ~$0.67 \pm 0.05$~ & ~$0.53 \pm 0.05$~ & ~$16.6 \pm 1.8$~ \\
\tECVCE & ~$0.72 \pm 0.06$~ & ~$0.14 \pm 0.02$~ & ~$53.0 \pm 1.2$~ \\
\end{tabular}
\caption{The axial-motion excitation $\Delta \bar{n}$ for a pair of \Be
ions transported in the same trapping well. Values for both axial modes of
motion (COM and stretch) are reported. The energy gain per trip is stated in
units of quanta where the COM frequency is 3.6~MHz and the stretch frequency is
6.2~MHz. Results from Ref.~\cite{blakestad2009a} are also given, though
only the COM mode excitation was investigated.}
\label{tab:heatingrates2ions} 
\end{table}

We expect (and observed) less excitation of the stretch mode relative to the COM
mode, for two reasons. First, the stretch mode frequency was higher than that of
the COM mode ($\omega_{\mathrm{STR}}=\sqrt{3} \omega_{\mathrm{COM}}$). Thus, the
filters on the rf and control potentials were more effective at suppressing noise
that could excite the stretch mode.  Second, a stretch mode can be excited only
by a differential force on the two ions, while the COM mode is excited by a force
common to both ions. Given the proximity of the ions to each other (a few
micrometers) compared to the distance of the ions to the trap electrodes, the
relative amplitude of differential forces acting on the ions are expected to be
less than common forces.

\section{Mitigation of Magnetic Field Fluctuations}\label{sec:Shield}

So far, we have discussed the suppression of undesired excitation of motional
degrees of freedom. We now discuss how magnetic-field fluctuations affecting
internal-state (qubit) coherence are suppressed in the X-junction trap array.

Decoherence of superpositions of the $\ket{2,-2}$ and $\ket{1,-1}$ qubit basis
states occurs both during transport and while the qubit is stationary. Previous
experiments demonstrated that junction transport contributed negligibly to
decoherence~\cite{blakestad2009a}. Magnetic field fluctuations form the dominant
contribution to qubit dephasing, yielding typical values (in this trap and
others) of less than $100~\mu$s~\cite{langer2005a}. Use of a
magnetic-field-insensitive qubit configuration enables extension of the coherence
time to approximately 10~s and can be used with some gate operations such as the
M{\o}lmer-S{\o}renson gate~\cite{molmer1999a}, but excludes implementation of
$\sigma_z \sigma_z$ gates~\cite{leibfried2003b, lee2005a, langer2005a}.

\begin{figure}
	\includegraphics{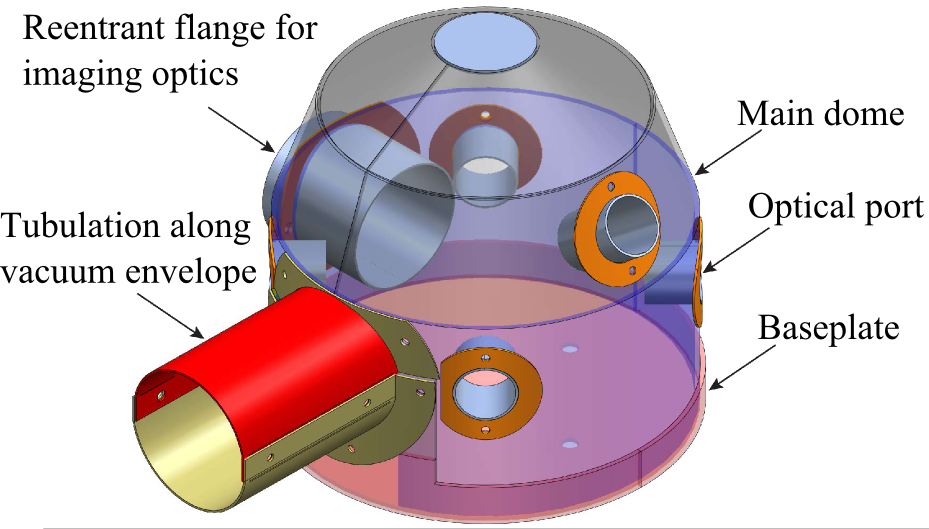}
	\caption{Mu-metal magnetic shield. The main dome enclosed both the trap and
	the magnetic-field coils. Cylindrical tubulation extended along a glass 
	vacuum envelope, which corresponds to the $\hat{z}$ direction at the trap.
	Reentrant flanges minimized field leakage around the imaging and optical access
	points. }
\label{fig:shield}
\end{figure}

To suppress the effects of magnetic field fluctuations, we enclosed the trap and
field coils inside a high-magnetic-susceptibility mu-metal shield and implemented
an active magnetic field stabilization system.  The shield
(Fig.~\ref{fig:shield}) was designed for compatibility with the existing trap
vacuum envelope and optical systems, and for ease of installation without the
need to lift the trap apparatus from the supporting table.  A cylindrical body
and approximately hemispherical dome were selected based on general guidelines
for magnetic shielding and manufacturing constraints.  The main body and
baseplate of the shield were constructed of $3.2$~mm thick, single-layer mu-metal
in order to provide maximum shielding of low-frequency magnetic field
fluctuations and to suppress magnetic saturation of the mu-metal. This latter
constraint arose because part of the field coils defining the quantization axis
of the qubits were located approximately 1~cm from the walls, where a calculated
field of $4 \times 10^{-3}$~T was expected for typical operating conditions.

Openings in the shield for optical components or laser beams were outfitted with a
reentrant flange fastened to the main body of the shield.  The flanges extended
both outwards and inwards in the shield in order to maximize flux-line
redirection.  A target length-to-diameter ratio of 5:1 guided design but was
typically not achieved due to geometric constraints arising from the exterior
dimensions of the vacuum envelope and the desire to position optical elements as
close as possible to the shield for maximum beamline stability.  These flanges
were designed to be modular, allowing for redesign and replacement if increased
shielding became necessary.  Magnetic continuity was achieved for all mating
flanges by use of internally threaded fasteners, producing a snug
contact fit.

The structure was designed to  provide a minimum 22~dB shielding of low-frequency
fields. This was confirmed by use of a pickup coil and detecting 60~Hz
fluctuations. The minimum suppression measured for the shield alone was $> 20$~dB
parallel to the $\hat{z}$ axis of the trap, and the direction of the largest
access opening in the shield. Shielding in excess of 60~dB  at 60~Hz was measured
in transverse directions.

The $1.3 \times 10^{-3}$~T quantization field is oriented
$45^{\circ}$ with respect to the vacuum envelope axis.  Initial measurements
identified current instability due to power-supply ripple as dominating measured
decoherence once the shield was installed. We implemented a custom
current-regulation system based on a proportional-integral-differential feedback
circuit, a series current-sense resistor, and a low-current field-effect
transistor. To minimize the effect of thermal drifts in the electronic circuit on
magnetic field stability, we selected special low-TC (thermal coefficient)
components and temperature-stabilized the enclosure.  The most critical
components were the gain and sense resistors; these were selected to be low-TC
metal foil resistors with less than $2$~ppm/K and less than $3$~ppm/K stability,
respectively. A four-terminal current-sense resistor was selected with
high-power-handling construction ($0.1~\Omega$ for the main coil and
$0.25~\Omega$ for the transverse shim coils). Similar care was taken to select
low-TC difference amplifiers for the input stage and a low-TC voltage reference.
All sense and feedback components were thermally sunk to an Al enclosure that was
thermally stabilized by use of Peltier coolers and a commercial temperature
controller with milliKelvin stability. Stabilization reduced current ripple from
$\sim 1$~mA to $\sim30~\mu$A on the main field-coil current of 1.2~A.  Net
magnetic field fluctuations due to current ripple at the location of the ions
were $\sim26$~nT.

Measurements of the dephasing time including both the magnetic shield and the
stabilization circuitry demonstrated extension of the qubit coherence to $1.41
\pm 0.09$~ms, more than 15$\times$ longer than that without shielding and
current stabilization, and sufficient for multiple transports before the qubit
dephased. A spin-echo pulse doubles the coherence time to $2.99 \pm 0.04$~ms,
indicating that slow shot-to-shot field fluctuations are small, and that
decoherence is dominated by fluctuations on a millisecond time scale.

\section{Mode Energy Exchange}\label{sec:Exchange}

The secular modes of the ions were constrained to change throughout the
transports, both in frequency and orientation. For most parts of the transport,
the splittings between the mode frequencies were sufficiently large and the
transport speed was sufficiently slow that modes changed adiabatically and energy
did not transfer between modes. However, at $\mathcal{C}$, the two principle axes
that lie in the $(x,z)$ plane were designed to have nearly degenerate secular
frequencies ($\omega'_x \approx \omega'_z \approx 2 \pi \times 5.7$~MHz), which
could lead to mode-mixing. The third mode along $\hat{y}$ had a significantly
higher frequency $\omega'_y = 2 \pi \times 11.3$~MHz, and would remain decoupled
from the $x$ and $z$ modes. Since the radial modes were only Doppler laser-cooled
before transport, $x/z$ mode mixing would increase the excitation of the axial
mode during transport. We employed two approaches that would minimize such axial
excitation. First, the duration during which the ion was at $\mathcal{C}$ could
be adjusted such to minimize the energy transfer between modes (by using a
duration that corresponded to a full cycle of the mixing process). Alternately, a
potential could be applied to various electrodes, which we will call the shim
potential, to sufficiently break the degeneracy (in practice, $\abs{\omega'_x -
\omega'_z} > 2 \pi \times 400$~kHz could be achieved) and suppress the mixing.
Both methods were effective and yielded similar transport excitation, though the
second approach was used for the results in Tables~\ref{tab:heatingrates}
and~\ref{tab:heatingrates2ions}.

However, in separate experiments, we explored a method for controlling energy
transfer between the motional modes of a single ion by using field shims near the
junction to tune $\omega'_x$ and $\omega'_z$ to near-degeneracy. Ideally, a
demonstration of the method would work as follows.  Prior to transport from
$\mathcal{E}$, the ion is cooled to the axial ground state $\ket{n_z = 0}$ along
$\hat{z}$ and prepared in Doppler-cooled thermal states in the transverse modes.
If the relative orientation of the modes remains stationary as the ion approaches
$\mathcal{C}$, the modes should not exchange energy, even if they become
degenerate. However, if the $x$ and $z$ mode directions diabatically (fast
compared to $1/\Delta \omega$) rotate $45^{\circ}$ to new directions given by
$x' = (x + z)/\sqrt{2}$ and $z' = (x - z)/\sqrt{2}$, we would expect the initial $x$
oscillation to project onto the new mode basis with half of the energy going into
each of the new modes. If $\Delta \omega' \equiv \omega'_x - \omega'_z \neq 0$,
the two oscillations would then begin acquiring a relative phase $\phi = \Delta
\omega' t$, where $t$ is the period spent at $\mathcal{C}$. By then quickly
transporting away from \tC\ towards $\mathcal{E}$ such that the mode axes rotate
diabatically by $-45^{\circ}$ back to their original orientation, the
oscillations would project back onto the original oscillator basis. If the wait
period is such that $\phi = \pi \times M$ (where $M$ is an integer), the motion
originally in the $x$ mode would project back into the same mode. If, however,
$\phi = \frac{\pi}{2} \times (2M-1)$, then the $x$-motion would project into the
$z$ mode; that is, the energy would exchange between $x$ and $z$ modes.

We demonstrated the basic features of this exchange as follows. To
tune the $\omega'_x$ and $\omega'_z$ close to degeneracy, an external shim
potential with an adjustable amplitude was applied. The shim potential
consisted of various contributions from 17 control electrodes near $\mathcal{C}$,
each multiplied by the overall scaling factor $A$. These contributions were
selected so that the net shim potential would primarily alter the frequency
splitting without significantly affecting other trapping parameters (such as the
position of the trapping minimum and the $y$ mode frequency).

\begin{figure}
	\includegraphics{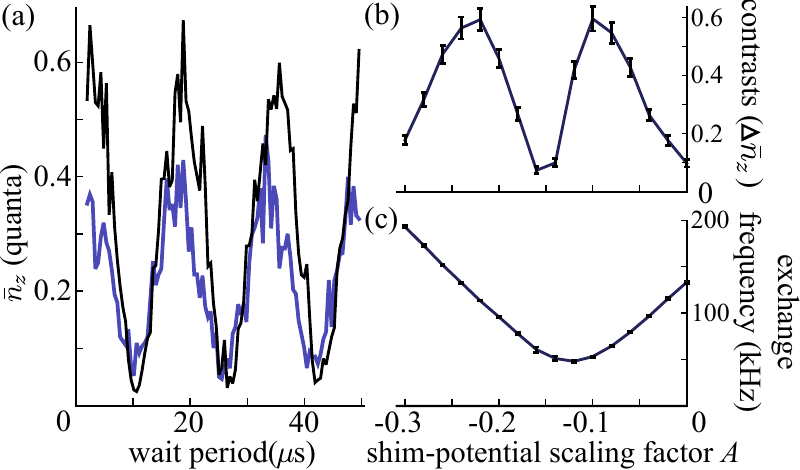}
	\caption{(a) Average motional excitation $(\bar{n}_z)$ in the axial mode after
	an \tECE\ transport, versus the duration at	$\mathcal{C}$ (wait period). The
	black trace indicates exchange of energy between the $z$ mode, prepared near
	the ground state (at $\mathcal{E}$), and the $x$ mode, prepared in a thermal
	state via Doppler cooling. The smaller blue trace represents the identical
	preparation, except the transport was performed twice. During the first
	transport, the wait period was set to maximize the energy transfer from the
	radial mode to the axial mode, followed by returning the ion to $\mathcal{E}$.
	After re-cooling the axial mode to the ground state, the round-trip transfer
	was repeated. The contrast was decreased (blue trace), indicating that
	less transverse mode energy was available for transfer to the $z$ mode and
	therefore indicating cooling of the $x$ mode. (b) and (c) The exchange contrast
	and exchange frequency (respectively), plotted versus the shim-potential
	scaling factor $A$.}
\label{fig:exchange}
\end{figure}

The black trace in Fig.~\ref{fig:exchange}(a) shows $\bar{n}_z$ for the axial
mode after \tECE\ transport, as the wait period at $\mathcal{C}$ was varied. We
derive $\bar{n}_z$ from sideband measurements, as described above, and assume a
thermal distribution. Although this assumption may not be strictly valid, it
should give a reasonable approximation for $\bar{n}_z < 1$. Oscillations between
an excited and near-ground state energy are visible. The projection process began
prior to the ion reaching $\mathcal{C}$ and transport was too slow for the
projection to be perfectly diabatic. Thus, the phase of the exchange oscillation
in Fig.~\ref{fig:exchange}(a) is not well determined and was observed to depend
on both the exchange frequency and the details of the approach to $\mathcal{C}$
(including speed and trajectory). In practice, it was difficult to maintain a
constant phase for more than a few minutes, as drifts in the potential, likely
caused by transient charge buildup and dissipation on the electrodes and also
pseudopotential amplitude changes, caused the exchange frequency to drift over
that time scale.

Figure~\ref{fig:exchange}(b) displays the oscillation contrast $\Delta \bar{n}_z
= \textrm{max}(\bar{n}_z) - \textrm{min}(\bar{n}_z)$ for various scaling factors,
$A$, of the shim potential, while Fig.~\ref{fig:exchange}(c) gives the frequency
of those oscillations versus the shim scaling factor. In separate experiments,
the two mode frequencies at $\mathcal{C}$ were measured as a function of $A$ by
driving excitations with an oscillatory potential applied to the control
electrodes. The difference between the two mode frequencies, $\Delta \omega' =
\abs{\omega'_x - \omega'_z}$, was observed to match the oscillation frequency of
the exchange process. Figure~\ref{fig:exchange}(c) suggests $\Delta \omega'$ is
high on the extreme ends of the $A$ range, while Fig.~\ref{fig:exchange}(b) shows
a reduction in contrast in these regions of high $\Delta \omega'$, likely due to
the reduction in diabaticity when $\Delta \omega'$ was large. This conclusion was
supported by the observation that the contrast decreased as the ion transport
speed was reduced. However, there was a maximum speed, above which contrast no
longer increased, because other sources of excitation began to obscure the
oscillatory signal.

From Fig.~\ref{fig:exchange}(b), we see that the fringe contrast was also
minimized for shim scaling factors near $A = -0.15$, where $\Delta \omega'$ was
small. This reduction in contrast can be explained as coinciding with the
condition where the initial mode orientation is identical to the rotated mode
orientation and thus the modes do not mix when projected, which is a condition
not necessarily related to $\Delta \omega'$. (We note that the $A$ value for
minimum exchange frequency does not match that for minimum contrast.)
Unfortunately, it was not possible to verify this, as we could not measure the
mode orientation at $\mathcal{C}$, due to lack of laser-beam access.

In the case where the energy from the $x$ mode was transferred into the $z$ mode,
the ion could be returned to $\mathcal{E}$ for a second round of ground-state
cooling of the $z$ mode. In the experiment, the exchange process in $\mathcal{C}$
was repeated and the results are shown as the blue trace in
Fig.~\ref{fig:exchange}(a), where a noticeable decrease in the ion's axial
excitation was observed compared to the first experiment without the second stage
of cooling. The small relative phase shift for the two traces in
Fig.~\ref{fig:exchange}(a) was due to the slow drift of $\Delta \omega'$ over the
several minutes required to take the two traces.

Ideally, all of the energy would be transferred from the $x$ mode into the $z$
mode, and the subsequent cooling of the $z$ mode would leave both modes in the ground
state, leading to no oscillation during the second trip into $\mathcal{C}$. In
practice, complete transfer was inhibited for two primary reasons. First, the
ions were not being transported fast enough to make a clean diabatic projection
of the motion onto the switched axes. Second, for complete energy transfer, the
projection should be onto axes rotated by $\pm 45^{\circ}$. Any other angles
would have resulted in incomplete transfer of energy. Attempts were made to
adjust additional shims in hopes of realizing configurations closer to $\pm
45^{\circ}$. However, as $\Delta \omega' \rightarrow 0$, it, again, becomes
difficult to predict the mode orientation with our idealized computer models and
we could not experimentally determine the mode axes in $\mathcal{C}$.

Nevertheless, we observed a clear and easily reproducible reduction in maximum
oscillation amplitude (Fig.~\ref{fig:exchange}(a)) from $\mathrm{max}(\bar{n}_z)
= 0.68 \pm 0.08$ to $\mathrm{max}(\bar{n}_z) = 0.40 \pm 0.05$ after the second
round of cooling, indicating the radial mode energy was being reduced. The use of
additional rounds of exchange followed by cooling reduced
$\mathrm{max}(\bar{n}_z)$ further, but after three or four exchange rounds, other
sources of excitation offset the energy reduction.

When optimized, this technique might be used to cool all modes of a single ion to
the ground state, while having the ability only to ground-state cool a single
mode, as for the laser beam configuration used here. A junction is not required;
all that is needed is a trap that can diabatically change the relevant mode
orientations by $\pm 45^{\circ}$, which could be possible in many trap
configurations.

\section{Conclusion}\label{sec:Conclusion}

In conclusion, we have demonstrated that transport through a two-dimensional trap
array incorporating a junction can be highly reliable and excite an ion's motion
by less than one quantum. This is a significant improvement over prior work with
junction arrays~\cite{hensinger2006a, blakestad2009a} and suggests the viability
of trap arrays incorporating junctions for use in large-scale ion-based quantum
information processing. In addition, we have implemented a mu-metal shield and
current stabilization to reduce qubit decoherence. We have also examined a
technique for transferring energy between motional modes.

DAC-update noise can be mitigated with the use of more appropriate filters such
as the Butterworth filters used here and/or faster DAC update rates. Noise on the
trapping rf potential can result in motional excitation at pseudopotential
barriers as described in Sec.~\ref{subsec:rfNoise}. The junction design criteria
in Ref.~\cite{amini2010a} included minimizing these barriers. However, the
results here show that the slope of the pseudopotential barrier is more important
than the barrier height for suppressing motional excitation, suggesting that
suppression of barrier height may not be a necessary constraint in future
designs.  Also, as observed here, with proper rf filtering, significant barrier
slopes can be tolerated without causing significant heating.

The technique for determining the waveforms described in this report can be
extended to incorporate multiple trapping wells by expanding
Eq.~(\ref{matrixConst}) to include multiple minima. Transport procedures such as
the ion exchange in Ref.~\cite{splatt2009a} are also amenable to these solving
techniques. Separating and combining of trapping wells requires consideration of
the potential's quartic term~\cite{home2006a}. Therefore, $\mathcal{P}$ in
Eq.~(\ref{derivVector}) could be expanded to include fourth-order derivatives.

With the use of multiple junctions, the techniques described here could help
provide a path toward transfer of information in a large-scale ion-based quantum
processor and enable an increased number of qubits in quantum algorithm
experiments. To do this, waveforms must be expanded to incorporate many trapping
wells. Also, it is likely that a sympathetic cooling ion species will need to be
co-trapped with the qubit ions to allow removal of the motional excitation from
electronic noise, multiple junction transports, and separating and re-combining
wells~\cite{home2009a, Hanneke2009a, jost2010a}. If sympathetic-cooling ions are
present, it may be advantageous to transport both ion species through a junction
in a single local trapping well. Since the pseudopotential (and micromotion) are
mass dependent, the qubit and cooling ions will experience different potentials,
which could lead to additional motional excitation. If such excitation is
excessive, it should still be possible to separate the ions into individual wells
by species and pass the different species through junctions separately, followed
by recombination.

We thank J.M. Amini, K.R. Brown, and J. Britton for contributions to the
apparatus and R. Bowler and J.P. Gaebler for comments on the manuscript. This
work was supported by IARPA, ONR, NSA, and the NIST Quantum Information program.
This is a publication of NIST and not subject to U.S. copyright.


\end{document}